%% file: main.tex
\documentclass[sigconf, nonacm]{acmart}

\newcommand\vldbdoi{XX.XX/XXX.XX}
\newcommand\vldbpages{XXX-XXX}
\newcommand\vldbvolume{16}
\newcommand\vldbissue{1}
\newcommand\vldbyear{2023}
\newcommand\vldbauthors{\authors}
\newcommand\vldbtitle{\shorttitle} 
\newcommand\vldbavailabilityurl{URL_TO_YOUR_ARTIFACTS}
\newcommand\vldbpagestyle{plain} 

\usepackage{balance}

\usepackage{mathtools}

\usepackage{subfigure}
\usepackage{multirow}
\usepackage{bm}
\usepackage{bbm}
\usepackage{dsfont}
\usepackage{color,soul}
\usepackage{url}
\usepackage{listings}
\usepackage{xcolor}
\usepackage{algorithmic}

\usepackage{natbib}

\usepackage[linesnumbered,vlined,ruled,noend]{algorithm2e}

\usepackage{threeparttable}

\usepackage{enumitem}
\usepackage{booktabs}
\usepackage{caption}
\usepackage{bm}

\usepackage{xspace}
\newcommand{\sys}{SiriusHelper\xspace}

\definecolor{codegray}{rgb}{0.5,0.5,0.5}
\begin{document}

\title{\sys: An LLM Agent-Based Operations Assistant for Big Data Platforms}



\author{Yu Shen*$^{\dagger}$,
Shiyang Liu*$^{\dagger}$,
Qihang He*$^{\dagger}$,
Yihang Cheng$^{\dagger}$,
Haining Xie$^{\dagger}$,
Zhiming He$^{\dagger}$,
Huahua Fan$^{\dagger}$,
Xianzhi Tan$^{\dagger}$,
Teng Ma$^{\dagger}$,
Shaoquan Zhang$^{\dagger}$,
Danqing Huang$^{\dagger}$,
Fan Jiang$^{\dagger}$,
Yang Li$^{\dagger}$,
Chongqing Zhao$^{\dagger}$,
Peng Chen$^{\dagger}$,
Jie Jiang$^{\dagger}$,
Bin Cui$^{\ddagger}$
}



\affiliation{
$^\dagger$TEG, Tencent Inc.\country{China}
$^\ddagger$School of Computer Science, Peking University\country{China}
}

\affiliation{
$^\dagger$\{willyushen, ethansyliu, cheesehe, yihangcheng, hainingxie, shimerhe, colinhhfan\}@tencent.com
\\
$^\dagger$\{linkatan, tatema, shawnqzhang, daisyqhuang, davyfjiang, thomasyngli, kendyzhao\}@tencent.com
\\
$^\dagger$\{pengchen, zeus\}@tencent.com $^\ddagger$bin.cui@pku.edu.cn
}

\renewcommand{\shortauthors}{Shen et al.}

\renewcommand{\authors}{Yu Shen, Shiyang Liu, Qihang He, Yihang Cheng, Haining Xie, Zhiming He, Huahua Fan, Xianzhi Tan, Teng Ma, Shaoquan Zhang, Danqing Huang, Fan Jiang, Yang Li, Chongqing Zhao, Peng Chen, Jie Jiang, Bin Cui}

\begin{abstract}
Big data platforms are widely used in modern enterprises, and an in-production intelligent assistant is increasingly important to help users quickly find actionable guidance and reduce operational burden. 
While recent LLM+RAG assistants provide a natural interface, they face practical challenges in real deployments: limited scenario coverage across both general consultation and domain-specific troubleshooting workflows, inefficient knowledge access due to inadequate multi-hop retrieval and flat knowledge organization, and high maintenance cost because escalated tickets are unstructured and hard to convert into assistant improvements and reusable SOPs.

In this paper, we present \sys, a deployed intelligent assistant for big data platforms. 
\sys serves as a unified online assistant that automatically identifies user intent and routes queries to the right handling path, including dedicated expert workflows for specialized scenarios (e.g., SQL execution diagnosis). 
To support complex troubleshooting, \sys combines a DeepSearch-driven mechanism with a priority-based hierarchical knowledge base to enable multi-hop retrieval without context overload, thus improving answer reliability and latency. 
To reduce expert overhead, \sys further introduces automated ticket understanding and SOP distillation: it diagnoses the assistant failure reason (e.g., missing knowledge or wrong routing) and extracts domain-specific SOPs to continuously enrich the knowledge base. 
Experiments and online deployment on Tencent Big Data platform show that \sys outperforms representative alternatives and reduces online ticket volume by 20.8\%.

\end{abstract}

\maketitle

\pagestyle{\vldbpagestyle}
\begingroup\small\noindent\raggedright\textbf{PVLDB Reference Format:}\\
\vldbauthors. \vldbtitle. PVLDB, \vldbvolume(\vldbissue): \vldbpages, \vldbyear.\\
\href{https://doi.org/\vldbdoi}{doi:\vldbdoi}
\endgroup
\begingroup
\renewcommand\thefootnote{}\footnote{\noindent
This work is licensed under the Creative Commons BY-NC-ND 4.0 International License. Visit \url{https://creativecommons.org/licenses/by-nc-nd/4.0/} to view a copy of this license. For any use beyond those covered by this license, obtain permission by emailing \href{mailto:info@vldb.org}{info@vldb.org}. Copyright is held by the owner/author(s). Publication rights licensed to the VLDB Endowment. \\
\raggedright Proceedings of the VLDB Endowment, Vol. \vldbvolume, No. \vldbissue\ %
ISSN 2150-8097. \\
\href{https://doi.org/\vldbdoi}{doi:\vldbdoi} \\
}\addtocounter{footnote}{-1}\endgroup


\input{intro}

\input{background}

\input{overview}
\input{design}

\input{knowledge}
\input{implementation}
\input{experiment}
\input{relatedwork}

\section{Conclusion}
\label{sec:conclusion}
In this paper, we presented \sys, an in-production intelligent assistant for big data platforms that supports both general consultation and domain-specific troubleshooting in a unified service. 
\sys combines an end-to-end workflow for non-specialized requests with automatic routing to specialized domain agents when deeper diagnosis is needed. 
For specialized scenarios, \sys leverages a hierarchical knowledge base that prioritizes authoritative and actionable sources, and adopts a DeepSearch-based execution loop with with planning and reflection to iteratively gather evidence while reducing retrieval cost and latency.
To reduce maintenance overhead, \sys further incorporates automated ticket understanding to identify failure causes and distill reusable SOPs from escalated cases. 
Evaluations on real-world queries from Tencent Big Data platform show that \sys achieves better answer quality than representative baselines, and reduces the volume of online tickets after deployment.

\balance

\bibliographystyle{ACM-Reference-Format}
\bibliography{reference}


\end{document}

%% file: intro.tex
\section{Introduction}
\let\thefootnote\relax\footnote{* Equal contribution.}

\label{sec:introduction}
Big data platforms have become the backbone of modern enterprises, powering data warehousing, real-time analytics, and large-scale ETL workflows~\cite{shahnawaz2025comprehensive, cuapusneanu2025reshaping}. 
In practice, platform teams must continuously respond to diverse user requests, ranging from routine usage questions (e.g., configuration, best practices) to highly specialized troubleshooting (e.g., SQL performance regression, Flink job failures). 
To support complex and evolving ecosystems, an \emph{in-production operations assistant} that can understand user intent, retrieve relevant knowledge, and guide users to actionable solutions is increasingly critical for improving operational efficiency and user experience.

Traditional AIOps tools~\cite{zhou2018bigroots, demirbaga2021autodiagn} and helpdesk workflows typically rely on handcrafted rules, static FAQs, or expert-maintained documentation. 
Although effective in narrow settings, these approaches struggle with the open-ended nature of user queries and fast evolution of big data systems. 
Meanwhile, recent advances in large language models (LLMs) in code understanding~\cite{crupi2025effectiveness,rahaman2024evaluating,huynh2025large} and reasoning~\cite{guo2025deepseek, chen2025towards, cheng2025empowering} make it possible to build a more natural interface for big data platforms.
Users can ask questions in plain language, and the assistant can generate guided suggestions by leveraging internal knowledge. 
As a result, many recent systems adopt LLMs together with Retrieval-Augmented Generation (RAG)~\cite{lewis2020retrieval} to retrieve relevant documents and historical tickets as evidence.
However, when applied to real-world big data systems, current LLM-based assistants still have several limitations:

\textbf{D1. Limited Scenario Coverage}.
Most existing systems either (i) focus on general consultation-style Q\&A with shallow workflows~\cite{singh2024panda} or (ii) target a specialized scenario (e.g., SQL diagnosis) with carefully engineered pipelines~\cite{cuiaetherlog, zhou2023d, pei2025flow}. 
In practice, a platform assistant must support both general questions and scenario-specific troubleshooting workflows.
It should also automatically route user queries, without requiring users to learn platform-specific patterns.

\textbf{D2. Inefficient Knowledge Access}.
User questions in big data platforms require evidence from multiple sources. 
A practical assistant needs to iteratively retrieve complementary evidence and then summarize the findings in an actionable answer. 
However, one-shot retrieval~\cite{singh2024panda, cuiaetherlog} is often insufficient, while naive multi-round retrieval~\cite{zhou2024llm, zhou2025gaussmaster} can easily cause context overload and degrade answer quality. 
Moreover, enterprise knowledge is typically stored in a flat collection (documents, FAQs, tickets), which makes it difficult to prioritize important sources (e.g., scenario SOPs), leading to unnecessary retrieval iterations and higher latency.

\textbf{D3. High Maintenance Cost}.
When an assistant fails to answer a user request, the case is usually escalated to a ticket. 
In principle, these tickets can be used to continuously improve the assistant, but it requires extensive manual effort. 
First, tickets must be analyzed to determine why the assistant failed and what kind of gap it exposed (e.g., missing knowledge, wrong routing), so that the right team can take action to improve the assistant and prevent similar escalations.
Second, for specialized troubleshooting scenarios (e.g., SQL/Flink diagnosis), the ticket resolution process contains actionable guidance.
However, this guidance is scattered across unstructured and multi-round conversations, making it difficult to manually consolidate into reusable SOPs.

To address these challenges, we present \sys, an operations assistant system for in-production big data systems. 
\sys is built as a unified online assistant that handles both open-ended user requests and deep, scenario-specific troubleshooting. 
For specialized scenarios (e.g., SQL execution diagnosis and Flink job diagnosis), \sys automatically routes queries into dedicated expert agents. 
Beyond online assistance, \sys incorporates automated ticket understanding and maintenance to identify assistant failure causes and continuously evolve the knowledge base with reduced expert overhead.

Our contributions are summarized as follows: \begin{enumerate} 
\item We present \sys, a \emph{deployed} operations assistant for big data platforms that lowers the barrier for end users. 
\sys automatically identifies user intent and routes queries to the appropriate workflows, without requiring users to master platform-specific expertise.

\item For specialized troubleshooting scenarios, we develop a DeepSearch-driven framework that combines a structured ``Plan-Retrieve-Filter'' loop with a priority-based hierarchical knowledge base. This design supports adaptive multi-hop retrieval with sufficient evidence, while avoiding context overload and hallucinations, and improving end-to-end latency.

\item To reduce expert overhead in knowledge maintenance, we implement an automated feedback mechanism from escalated tickets: \sys analyzes escalated tickets to diagnose the failure reason (e.g., missing knowledge, wrong routing), and distills workflow-specific SOPs from the categorized tickets to enrich the knowledge base.

\item Finally, we evaluate \sys on real-world data from Tencent Big Data platform, covering both user tickets and technical consultations. Empirical results show that \sys achieves superior performance compared with representative alternatives, and reduces online ticket volume by 20.8\% after deployment. 
\end{enumerate}

\underline{\textit{Overview}}. 
The rest of this paper is organized as follows. 
Section~\ref{sec:background} introduces the problem background. Section~\ref{sec:sys_overview} provides an overview of the general pipeline. Section~\ref{sec:system_design} presents the design for specialized troubleshooting scenarios, including the hierarchical knowledge base and the DeepSearch-based workflows. Section~\ref{sec:knowledge} describes how knowledge is automatically maintained and improved over time, including ticket understanding and SOP distillation.
Section~\ref{sec:exp} presents experimental evaluations and online impact. 
Finally, Section~\ref{sec:related} reviews related work and Section~\ref{sec:conclusion} concludes the paper.

%% file: background.tex
\section{Preliminary}
\label{sec:background}

\subsection{Problem Definition}
Let the incoming user request be denoted by $u$.
An in-production AIOps assistant aims to generate an actionable response for $u$ with low latency by leveraging enterprise resources, including heterogeneous knowledge bases as well as operational tools.
Since user requests are often ambiguous or incomplete, the assistant may optionally rewrite the request into a clearer query $q$ for subsequent retrieval and reasoning.

A critical challenge is to ground the response on reliable evidence.
We denote the set of knowledge sources as $\mathcal{K}=\{\mathcal{K}^1, \ldots, \mathcal{K}^L\}$ (e.g., SOPs, FAQs, technical documents, and historical tickets), which differ in coverage and reliability.
The assistant must retrieve relevant evidence under latency and context-budget constraints.

\subsection{Retrieval-Augmented Generation (RAG)}
Retrieval-Augmented Generation (RAG) augments LLM generation with evidence retrieved from external knowledge bases~\cite{lewis2020retrieval, fan2024survey}.
Given a query $q$, the retriever returns a small set of evidence passages $E$ from $\mathcal{K}$, and the LLM generates the final response conditioned on $q$ and $E$.
In enterprise settings, retrieval is commonly implemented as a multi-stage pipeline, e.g., coarse retrieval followed by reranking, to balance latency and quality.

\begin{algorithm}[t]
  \caption{Two-stage RAG process.}
  \label{algo:rag_retrieval}
  \begin{algorithmic}[1]
  \REQUIRE Query $q$, knowledge base $\mathcal{K}$, coarse retrieval size $N$, evidence size $k$.
  \ENSURE Evidence set $E$.
  \STATE $C \leftarrow \textsc{CoarseRetrieve}(q,\ \mathcal{K},\ N)$
  \STATE $C' \leftarrow \textsc{Rerank}(q,\ C)$
  \STATE $E \leftarrow \textsc{SelectTopK}(C',\ k)$
  \STATE $E \leftarrow \textsc{Deduplicate}(E)$
  \RETURN $E$
\end{algorithmic}
\end{algorithm}

Algorithm~\ref{algo:rag_retrieval} illustrates a standard two-stage RAG retrieval procedure.
This architecture enables knowledge updates by changing the external database, which can mitigate hallucinations without retraining the model.
Despite its effectiveness, conventional RAG is typically implemented as a static single-turn pipeline: the query is generated once, retrieval is performed once, and the answer is produced without further interaction.
This limitation becomes apparent in complex multi-step reasoning tasks, which often require iterative refinement and adaptive exploration with different queries or across distinct information sources~\cite{shao2023enhancing}.
As a result, single-turn RAG may underperform on tasks that require sustained reasoning over multiple rounds.

\begin{figure*}[t]
	\centering
	\scalebox{0.95}[.95] {
	\includegraphics[width=1\linewidth]{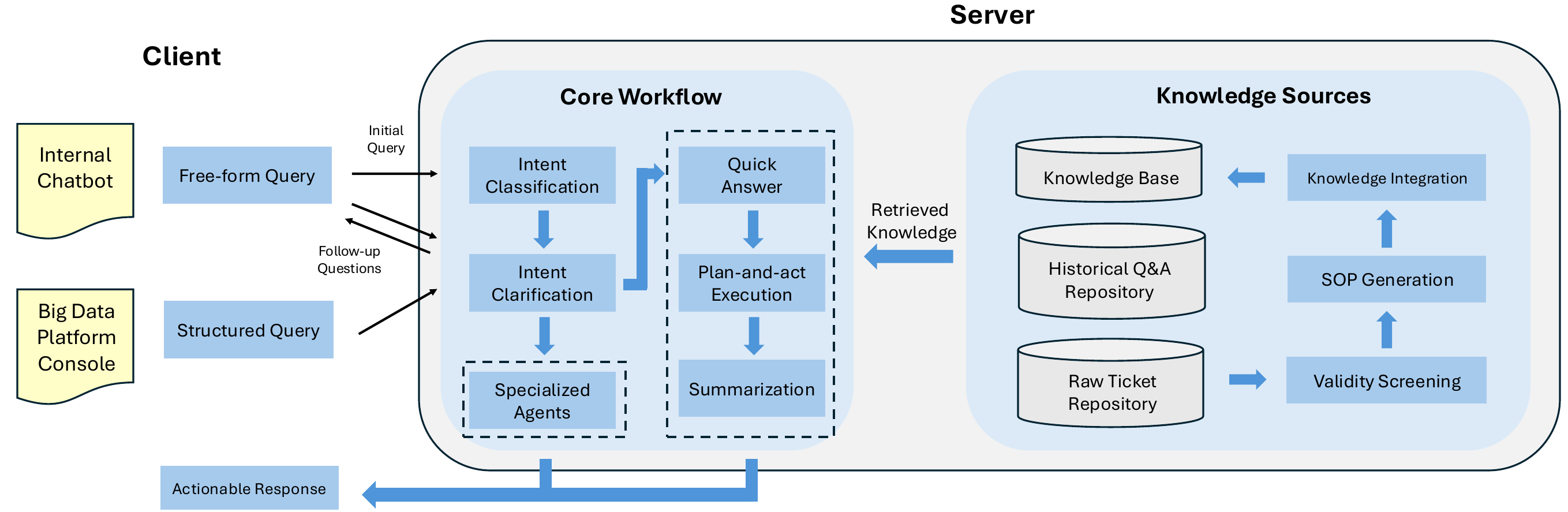}
         }
	\caption{\sys system overview.}
    \label{fig:system_overview}
\end{figure*}

\subsection{Evolution to DeepSearch}

The evolution from conventional RAG to DeepSearch reflects a shift toward more iterative and goal-driven information seeking~\cite{shao2023enhancing, xi2025survey, singh2025agentic}. 
Building on this direction, recent work has proposed DeepSearch agents, which further increase autonomy through:

\begin{enumerate}
    \item \textbf{Multi-turn search}: The agent iteratively refines queries, retrieves evidence, and summarizes intermediate findings.
    \item \textbf{Multi-source retrieval}: The agent gathers information from different sources, such as web pages and private databases.
\end{enumerate}

Under this agentic setting, search is defined as an active loop: the agent repeats retrieval and synthesis until it has collected enough evidence to answer the question. 
Compared with single-turn RAG, DeepSearch emphasizes proactive and structured exploration. Representative systems include WebDancer~\cite{wu2025webdancer}, OpenAI DeepResearch~\cite{openai2025gpt52}, and Gemini DeepResearch~\cite{google2025geminideepresearch}, which generate reports by iteratively collecting and organizing evidence from multiple sources.

Overall, RAG provides a foundation for grounding LLM outputs in external knowledge, while DeepSearch extends RAG into an iterative framework for multi-step information gathering and reasoning.

%% file: overview.tex
\section{System Overview} 
\label{sec:sys_overview}
In this section, we first present the overall architecture of \sys, and then describe the general workflow used to handle non-specialized requests.

\subsection{Overview}
\label{sec:overview}
Figure~\ref{fig:system_overview} presents the system architecture of \sys. 
The system follows a client-server design to support multiple entry points with centralized reasoning. 
On the client side, \sys currently provides two primary interfaces:
(i) a platform console, which directly collect runtime context (e.g., task ID, SQL text, and error logs) from the execution page, and
(ii) a chatbot embedded in the internal messaging tool, which lowers the barrier for initiating requests via free-form natural language.

On the server side, \sys runs as an independent service responsible for request understanding, task routing, and response generation.
Structured requests from the console can be directly sent to the router.
In contrast, chatbot requests typically require additional clarification, after which they enter the same backend execution pipeline.
The server maintains three knowledge sources: a curated knowledge base (e.g., domain documents), a historical Q\&A repository for fast resolution, and a ticket repository containing historical issue reports and their resolutions.
These components enable \sys to provide consistent support across multiple entries and lightweight interactions for end users.

\subsection{General Workflow}
\label{sec:general_workflow}
\sys follows a general workflow to handle user requests in a uniform manner, especially for chatbot interactions where the input context is often incomplete.
The workflow consists of the following steps:

\textbf{Stage 1: Intent Classification}. \sys first determines whether the user message is actionable (e.g., consultation or troubleshooting) or non-actionable (e.g., chit-chat). For non-actionable queries, \sys responds with a brief and polite refusal.

\textbf{Stage 2: Intent Clarification and Routing}. For actionable requests, \sys performs intent clarification by rewriting the question into a structured form.
When necessary, \sys asks follow-up questions to collect missing fields. 
Based on the clarified intent, the request is routed to the corresponding agent.

\textbf{Stage 3: Quick Answer}. 
For non-specialized tasks, \sys first searches the ticket repository to identify previously solved cases with high similarity, and it returns a concise answer when a confident match is found. If no similar ticket is retrieved, \sys checks whether the clarified request is sufficiently simple and can be answered directly with the available context.

\textbf{Stage 4: Plan-and-act Execution}. 
If the request still cannot be resolved, \sys generates a plan and executes it through RAG and tool calls, collecting evidence until it can form a grounded solution.

\textbf{Stage 5: Summarization}. Finally, \sys summarizes the evidence and produces a clear response, including key conclusions and actionable next steps.

Note that console-based requests skip the classification and clarification stage, as the interface already enforces structured inputs, but they share the same routing and backend execution pipeline.
When the request is routed to a specialized domain agent, the routing stage invokes a domain-specific execution workflow to complete troubleshooting under a particular system context. 
We detail this specialized agent workflow in Section~\ref{sec:system_design}.

%% file: design.tex
\begin{figure*}[thb]
	\centering
	\scalebox{0.95}[.95] {
	\includegraphics[width=1\linewidth]{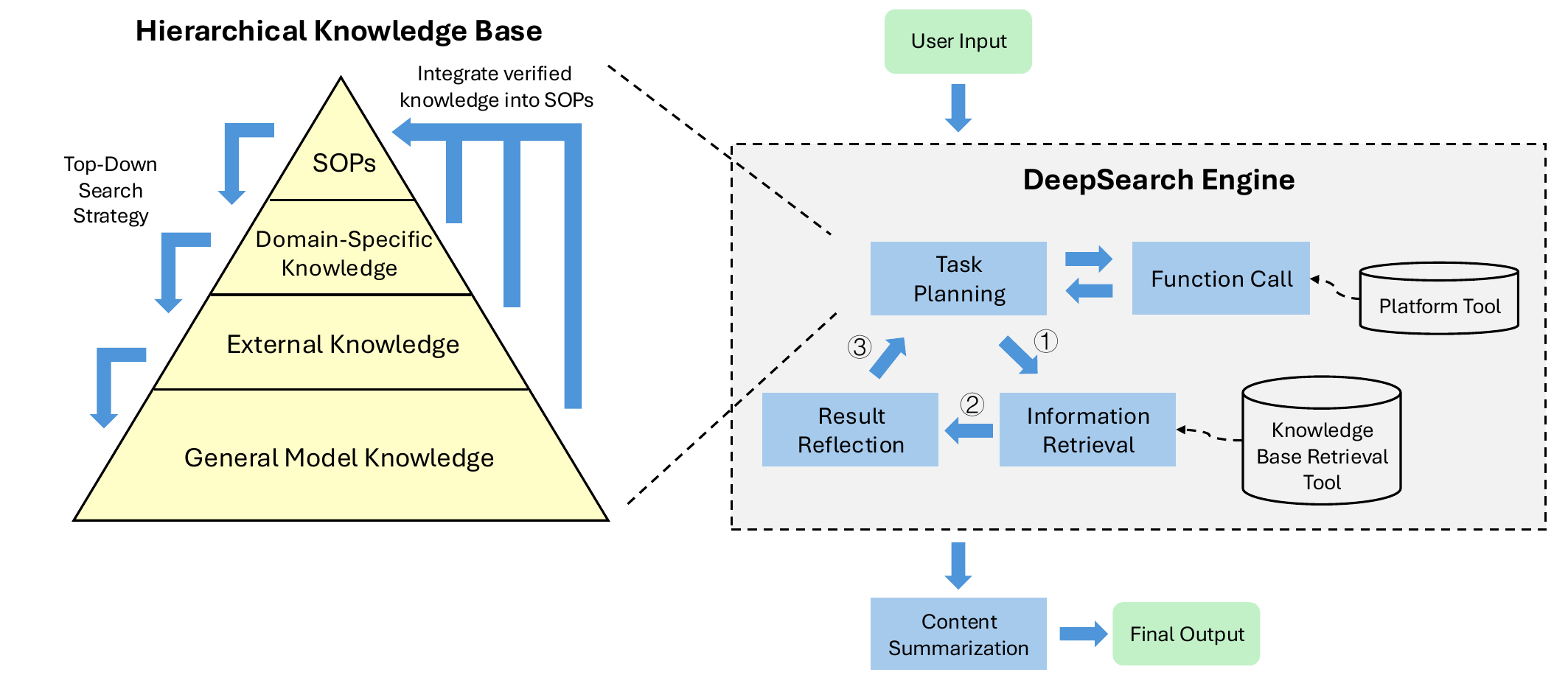}
         }
	\caption{Specialized agent workflow in \sys.}
    \label{fig:workflow}
\end{figure*}

\section{Specialized Agent Design}

\label{sec:system_design}
In this section, we describe \sys’s specialized agent workflow and the key components.

\subsection{Agent Overview}
The overall agent workflow is illustrated in Figure~\ref{fig:workflow}.
In the general workflow, as \sys has already collected necessary clarifications when needed, the specialized agent is triggered with a complete and structured problem description.

Based on the structured input, \sys enters the \textbf{DeepSearch engine}, in which an iterative reasoning loop consists of three steps: 
(1) \textbf{task planning} (Step 1), where \sys maintains the current solving state (e.g., structured intent and accumulated evidence) and decides the next action, either invoking a platform tool (e.g., to fetch runtime metrics/logs) or performing knowledge retrieval. 
If retrieval is selected, \sys further chooses an appropriate level in the hierarchical knowledge base and formulates a targeted query; 
(2) \textbf{information retrieval} (Step 2), where \sys queries a retrieval tool to collect relevant evidence; 
(3) \textbf{result reflection} (Step 3), where intermediate results are checked for completeness and consistency, and the update state is then fed back to the task planning module to start the next iteration. 
After the loop converges, the system conducts \textbf{content summarization} and generates the final output.
The agent workflow pseudo-code is provided in Algorithm~\ref{algo:workflow}.

To support reliable retrieval, the knowledge base is organized as a hierarchical pyramid, ranging from \textbf{general model knowledge} and \textbf{external knowledge} to \textbf{domain-specific knowledge} and finalized \textbf{SOPs}. 
The system follows a top-down search strategy and continuously integrates verified knowledge into SOPs for reuse in future tasks.

\begin{algorithm}[t]
  \caption{Pseudo-code of specialized agent workflow.}
  \label{algo:workflow}
  \begin{algorithmic}[1]
  \REQUIRE Structured request $q$, hierarchical knowledge base $\mathcal{K}=\{\mathcal{K}_{\text{SOP}},\mathcal{K}_{\text{INT}},\mathcal{K}_{\text{WEB}},\mathcal{K}_{\text{LLM}}\}$, tool set $\mathcal{F}$, max iterations $T$.
  \ENSURE Final response $a$.
  \STATE Initialize solving state $S \leftarrow \{ \text{intent}=q,\ \text{evidence}=\varnothing\}$ 
  \FOR{$t=1$ \TO $T$}
    \STATE \# $\textit{act}\in\{\textsc{Tool},\textsc{Retrieve}\}$, $u$: tool call (name + args), $l$: knowledge level, $r$: retrieval query.
    \STATE $(\textit{ans\_ready}, \textit{act}, u, l, r) \leftarrow \textsc{Planner}(S, \mathcal{K}, \mathcal{F})$
    \IF{$\textit{ans\_ready}=\textbf{true}$}
      \STATE \textbf{break}
    \ENDIF
    \IF{$\textit{act}=\textsc{Tool}$}
      \STATE $o \leftarrow \textsc{ToolExecutor}(\textit{act}, \mathcal{F})$
      \STATE $S.\text{evidence} = S.\text{evidence} \cup \left(\textit{act}, o\right)$
      \ELSIF{$\textit{act}=\textsc{Retrieve}$}
      \STATE $C \leftarrow \textsc{Retriever}(r, \mathcal{K}_{l})$
      \STATE $C' \leftarrow \textsc{Filter}(r, C)$
      \STATE $S.\text{evidence} = S.\text{evidence} \cup \left(act, C'\right)$
    \ENDIF
  \ENDFOR
  \STATE $a \leftarrow \textsc{Summarize}(S)$
  \RETURN $a$
\end{algorithmic}
\end{algorithm}

\subsection{Standard Operating Procedure (SOP)}
\label{sec:sop}
In this paper, we treat SOPs as the first-priority knowledge source for automated troubleshooting in specialized agents. 
An SOP describes an operator-validated diagnostic workflow for a recurring failure pattern. 
Compared with generic troubleshooting documents, SOPs are more actionable and structured. 
They specify the expected investigation path and the concrete operations at each stage, which reduces the search space and helps LLMs behave more stably in domain-specific troubleshooting scenarios.

In \sys, each SOP is stored as a structured record with the following fields: 
(1) \textbf{Problem description}, which captures observable symptom signatures, such as key error codes or representative log patterns, as well as concise natural-language descriptions summarized from user questions. 
(2) \textbf{Root cause}, which describes the underlying failure source at an actionable level. 
(3) \textbf{Investigation steps}, which are organized as an ordered workflow. 
Each step specifies (i) the target to inspect, (ii) the action to perform, and (iii) the expected observations that decide whether to continue to the next step or terminate when the root cause is confirmed.
(4) \textbf{Resolution steps}, which provide concrete resolution operations to address the confirmed root cause. 
In addition, each SOP includes provenance metadata indicating whether it is manually authored or automatically distilled from a specific historical ticket by our pipeline. 
This provenance is surfaced during answer generation to support traceability, enabling engineers to trace back and revise the SOP when necessary.

\begin{lstlisting}[label={lst: sop}]
problem description: key error codes/descriptions

root cause 1: ...

- investigation steps: 
    (1) [target]: code modules, logs, data sources, 
                  permissions, resource quotas, etc.
        [action]: obtain <field> info of <table>, 
                  check <package> version, 
                  verify <file> permissions, etc.
        [observations]: A: proceed to step x,
                       B: root cause 1 confirmed,
                       ...
    (2) ...
    
- resolution steps:
    (1) ...
    (2) ...
    
root cause 2: ...
\end{lstlisting}

For indexing and retrieval, SOPs are integrated into the knowledge base using the problem description as the retrieval key, while the entire SOP record (problem description, root cause, investigation and resolution steps) is stored as the associated value. 
This design aligns retrieval with the primary observation (e.g., error messages and question description) available at inference time.

\subsection{Hierarchical Knowledge Base}
\label{sec:knowledge_base}
In \sys, the general knowledge sources are treated as a flat collection during retrieval, without an explicit priority order. 
For specialized agents, we additionally construct a structured, four-level hierarchical knowledge base. 
This hierarchy is manually designed as a pyramid according to specificity, authority, and actionability, providing an explicit retrieval path that prioritizes more reliable and directly executable knowledge whenever possible.

\textbf{Level 1: Standard Operating Procedures (SOPs)}.
This level contains the most actionable knowledge, typically represented as key-value entries in the form of ⟨error code/description, root cause and resolution⟩. 
The SOPs are sourced from (i) authoritative procedures manually curated by domain experts and (ii) solutions automatically distilled from historical ticket records. 
The goal of this level is to provide the fastest and most reliable resolution for recurring issues.

\textbf{Level 2: Internal Professional Knowledge Base.}
This level stores organization-specific expertise, including component documentation, internal FAQs, and architecture descriptions of key system components (e.g., SQL engines and file systems).
Knowledge is organized as ⟨question title, content passage⟩ pairs, serving as a contextual and in-depth complement to SOPs.
When no direct SOP applies, the system retrieves relevant passages from this layer to support diagnosis and resolution.

\textbf{Level 3: Open Web Knowledge}.
When internal knowledge is insufficient, \sys activates this layer by leveraging integrated commercial search engines and web crawling APIs to collect public technical documentation, community discussions, and other online resources. 
This layer provides broader background information and alternative hypotheses for troubleshooting.

\textbf{Level 4: Model General Knowledge}.
As the final fallback, this level does not rely on external storage; instead, it utilizes the large language model’s general technical knowledge acquired during pretraining and its reasoning capability. When none of the above sources yields useful evidence, the system generates responses based on this general knowledge.

\textbf{Continual self-enrichment}.
The knowledge base is designed to evolve over time. As the system continuously processes online tickets, it automatically learns from resolution traces and internal documents to summarize new SOP entries, thereby enriching the top layer. This progressive refinement shifts an increasing portion of queries to higher layers, enabling faster resolution and improving overall system efficiency.

\subsection{DeepSearch Engine}
\label{sec:deepsearch}
The DeepSearch engine is the core component that enables intelligent, iterative analysis in our framework. 
It drives a closed-loop process by coordinating a planner, a retriever, and a filter to repeatedly perform planning, retrieval, and selection until sufficient evidence is gathered.

\textbf{Planner}. 
The planner serves as the decision maker. Based on the current analysis state and accumulated evidence, it first assesses whether the evidence is sufficient to answer the original question.
If so, it terminates the loop and triggers final summarization. Otherwise, it determines the next action to acquire information, which can be either (i) \emph{function call} by choosing a platform tool (e.g., fetching job logs/metrics or configuration) to collect runtime evidence, or (ii) \emph{knowledge retrieval} by selecting an appropriate scope in the hierarchical knowledge base (e.g., staying in the current layer or switching to another layer) and generating a focused retrieval query that targets the most salient missing information (e.g., “how to resolve a mismatch in the number of aliases for \texttt{LATERAL VIEW EXPLODE}”).

\textbf{Retriever}. 
Guided by the planner’s selected scope and query, the retriever searches the specified knowledge source and returns candidate passages that are potentially relevant to the current query.

\textbf{Filter}. 
The filter evaluates the retrieved candidates against the planner’s query and intent, discarding irrelevant or noisy results and retaining only evidence that is consistent with the current investigation. The filtered evidence is then fed back to the planner to initiate the next iteration.

By iteratively coordinating these three modules, DeepSearch mimics an expert’s process of proposing hypotheses, gathering evidence, and validating findings, thereby improving the depth and robustness of troubleshooting in specialized agents. 
Note that, unlike conventional DeepSearch pipelines~\cite{xi2025survey}, we explicitly introduce a \textbf{filter} stage in each iteration to remove irrelevant retrieval results before they enter the reasoning context. 
Without this step, ambiguous or misleading solutions are more likely to be accumulated across iterations, which increases the risk of hallucinated reasoning and incorrect root-cause attribution in the final diagnosis.

%% file: knowledge.tex
\section{Knowledge Maintenance}
\label{sec:knowledge}

To keep \sys effective under evolving platforms, we design a knowledge maintenance module that continuously enriches support knowledge from daily tickets. 
Rather than relying solely on manual curation, the module builds a closed loop over ticket conversations: it (i) classifies tickets via LLM-based categorization, (ii) performs confidence-aware routing to reduce manual attribution cost, and (iii) converts category-specific tickets into actionable SOPs, which are then integrated back into the hierarchical knowledge base.

\subsection{Ticket Categorization}
A ticket is a structured record of dialogue initiated to resolve a concrete issue, including the user’s problem description, multi-turn interactions, and the final handling outcome. 
To enable ownership attribution and knowledge maintenance, \sys leverages an LLM to automatically annotate each ticket with a set of categorical and semantic labels, including \textbf{system}, \textbf{module}, \textbf{request type} (consultation vs. troubleshooting), \textbf{ticket summary}, \textbf{keywords}, and \textbf{final action} (e.g., debugging, permission granting, or handoff). 
These labels provide a unified view of heterogeneous tickets and serve as the basis for subsequent probabilistic assignment and category-aware SOP extraction.

\subsection{Probabilistic Assignment} 
In production systems, each ticket captures a case where \sys does not effectively resolve a user issue and requires further follow-up. 
Therefore, for each ticket, we need to identify the likely cause and route it to the right owner for improvement. 
We maintain a list of nine common causes (e.g., missing knowledge). 
To avoid manually attributing every ticket, we design a posterior-based probabilistic assignment process that automatically assigns high-frequency cases and leaves the rest for human review.

We treat the ticket’s \textbf{final actions} as categorical features (e.g., data onboarding, troubleshooting, permission granting).
A single ticket may include multiple features, and we assume these features are conditionally independent given the underlying cause.
We estimate (i) the prior probability $P(c)$ for each cause $c$, and (ii) the conditional probability $P(f\mid c)$ as the empirical frequency of action feature $f$ among tickets attributed to cause $c$. For a ticket represented by a feature set $F=\{f_1,\ldots,f_n\}$, we compute the posterior by Bayes’ rule:
$$
P(c\mid F) \propto P(c)\prod_{f\in F} P(f\mid c).
$$

For example, consider a specific cause $c_1$ with prior $P(c_1)=0.3$, and a ticket with $F=\{f_1,f_2,f_3\}$ where $P(f_1\mid c_1)=0.8$, $P(f_2\mid c_1)=0.3$, and $P(f_3\mid c_1)=0.25$. The unnormalized posterior score is $0.8\times 0.3\times 0.25\times 0.3$. Since $P(F)$ is constant across causes, we normalize scores to compute the posterior. 
If the largest posterior exceeds $0.8$, we automatically assign the ticket to that cause and route it to the corresponding owner, otherwise we defer to manual attribution.
This strategy captures high-frequency patterns and substantially reduces per-ticket diagnosis cost.

\begin{figure*}[t!]
	\centering
	\scalebox{0.95}[.95] {
	\includegraphics[width=1\linewidth]{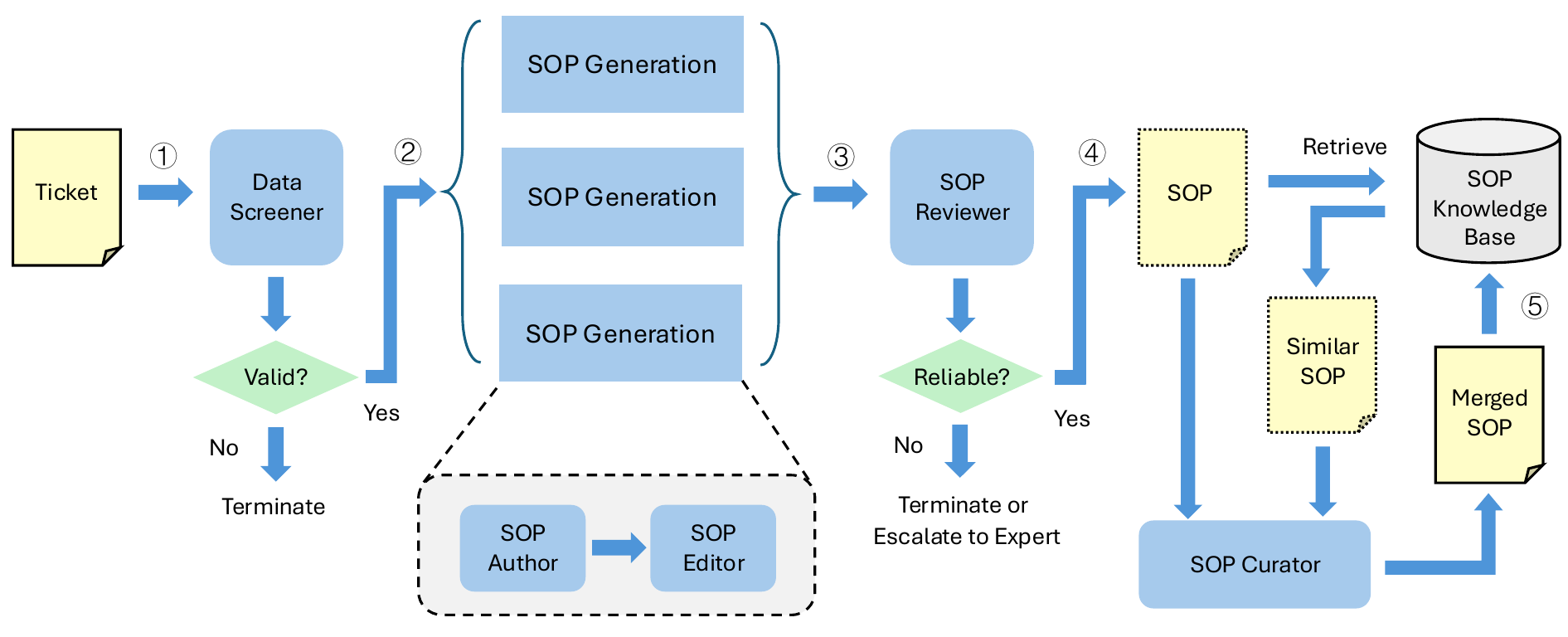}
         }
	\caption{SOP extraction workflow.}
    \label{fig:knowledge self-enrichment workflow}
\end{figure*}

\subsection{SOP Extraction}
\label{sec:knowledge_self_enrichment}
To mitigate the cost and staleness of manually curated knowledge, \sys incorporates a pipeline that automatically extracts SOPs from historical tickets.
It takes as input the tickets that have been assigned to a specific category (i.e., system and module) by ticket categorization, ensuring that SOP generation is performed within the correct domain context.
Figure~\ref{fig:knowledge self-enrichment workflow} illustrates the entire process, which is implemented as a multi-agent LLM workflow consisting of three stages: data validity screening, structured SOP generation, and knowledge base integration.
This design ensures that only high-quality, operator-actionable SOPs are admitted into the system. 
The workflow pseudo-code is shown in Algorithm~\ref{algo:knowledge_self-enrichment}.

\begin{algorithm}[t]
  \caption{Workflow pseudo-code of SOP extraction.}
  \label{algo:knowledge_self-enrichment}
  \begin{algorithmic}[1]
  \REQUIRE Historical ticket document $D$, SOP knowledge base $\mathcal{K}_{\text{SOP}}$, repeating SOP generation times $N$.
  \ENSURE updated SOP knowledge base $\mathcal{K}_{\text{SOP}}$.
  \STATE $is\_valid \leftarrow \textsc{DataScreener}(D)$
  \IF{$\textit{is\_valid}=\TRUE$}
    \STATE SOP candidate set $S \leftarrow \varnothing$
    \FOR{$t=1$ \TO $N$}
        \STATE $s \leftarrow \textsc{SOPAuthor}(D)$
        \STATE $s \leftarrow \textsc{SOPEditor}(D, s)$
        \STATE $S \leftarrow S \cup \{s\}$
    \ENDFOR
    \STATE $is\_reliable, s^* \leftarrow \textsc{SOPReviewer}(S)$
    \IF{$\textit{is\_reliable}=\TRUE$}
        \STATE $similar\_SOPs \leftarrow \textsc{Retriever}(s^*, \mathcal{K}_{\text{SOP}})$
        \STATE $has\_merged \leftarrow \FALSE$
        \FOR{$s$ \textbf{in} $similar\_SOPs$}
            \STATE $has\_merged, s' \leftarrow \textsc{SOPCurator}(s, s^*)$
            \IF{$\textit{has\_merged}=\TRUE$}
            \STATE \# Replace $s$ by $s'$ in the $\mathcal{K}_{\text{SOP}}$.
            \STATE $\mathcal{K}_{\text{SOP}} \leftarrow \textsc{Replace}(s, s', \mathcal{K}_{\text{SOP}})$
            \STATE \textbf{break}
            \ENDIF
        \ENDFOR
        \IF{$\textit{has\_merged}=\FALSE$}
        \STATE \# Add new entry $s^*$ to the $\mathcal{K}_{\text{SOP}}$.
        \STATE $\mathcal{K}_{\text{SOP}} \leftarrow \textsc{Add}(s^*, \mathcal{K}_{\text{SOP}})$
        \ENDIF
    \ENDIF
  \ENDIF
  \RETURN $\mathcal{K}_{\text{SOP}}$
\end{algorithmic}
\end{algorithm}

\textbf{Stage 1: Data Validity Screening}.
The first stage improves downstream reliability by filtering out noisy, incomplete, or non-actionable tickets before SOP construction. 
Given raw historical dialogues, an agent named \textbf{Data Screener} performs a binary validity check via a LLM-based semantic audit. 
It determines whether the core issue in the ticket is still unresolved, has only been addressed with a temporary workaround due to a known bug, or is intermittent.
Invalid tickets are discarded, and only valid tickets are passed to the SOP generation stage.

\textbf{Stage 2: Structured SOP Generation}.
This stage generates structured SOPs and reduces hallucination through an iterative ``generate-verify'' loop with stability evaluation.
Three agents collaborate: 
(i) \textbf{SOP Author} extracts the diagnostic logic from the conversation and produces an SOP draft, as shown in Section~\ref{lst: sop}. The draft may include multiple candidate root causes. 
(ii) \textbf{SOP Editor} revises the draft based on the original conversation by removing redundant branches, clarifying ambiguous steps, and keeping only the most defensible root cause.
(iii) \textbf{SOP Reviewer} evaluates multiple drafts generated from the same ticket to check cross-version consistency. It outputs a stability score and selects the final SOP from the candidate drafts. Only SOPs that meet a predefined stability threshold are accepted; otherwise, the outputs are rejected and escalated for expert review.

\textbf{Stage 3: Knowledge base Integration}.
This stage integrates verified SOPs into the SOP knowledge base while avoiding duplicates.
For each accepted SOP, we retrieve existing SOPs using its problem description to find potential near-duplicates.
A dedicated agent, \textbf{SOP Curator}, decides whether the new SOP should be merged with an existing one and produces a canonical entry. 
If a similar SOP exists, the curator merges them by combining their problem descriptions into a richer symptom signature. 
If the root causes are also similar, it merges the root-cause descriptions and consolidates the corresponding investigation and resolution steps. Otherwise, it adds the new root cause as an additional branch under the merged SOP. The merged SOP is stored back into the knowledge base, supporting continual updates and reducing redundancy.
A case study is provided in Section~\ref{sec:sop_case_study}.

%% file: implementation.tex
\section{System Implementation}
\label{sec:implementation}
In this section, we present the implementation details of \sys. 
We first introduce the system interface, and then describe the supporting modules that enable the end-to-end workflow (e.g., intent clarification and retrieval-based reasoning). 
Finally, we provide representative prompt templates used by the key modules.

\subsection{Product Interface}
\begin{figure}[t]
    \centering
    \includegraphics[width=0.9\linewidth]{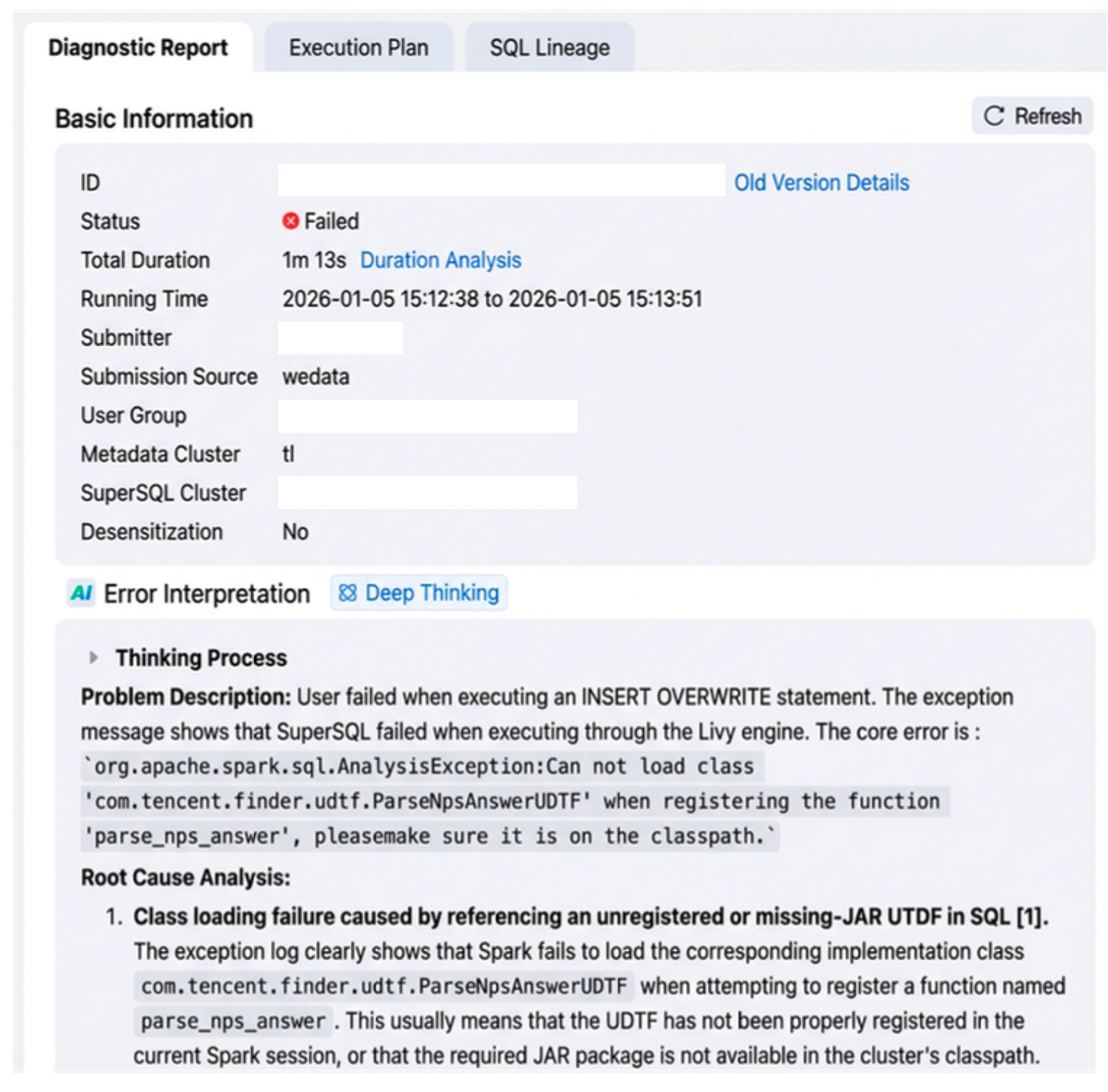}
    \caption{Product interface I: In-console diagnosis.}
  \label{fig:in_console_diagnosis}
\end{figure}

\begin{figure}[t]
    \centering
    \includegraphics[width=0.9\linewidth]{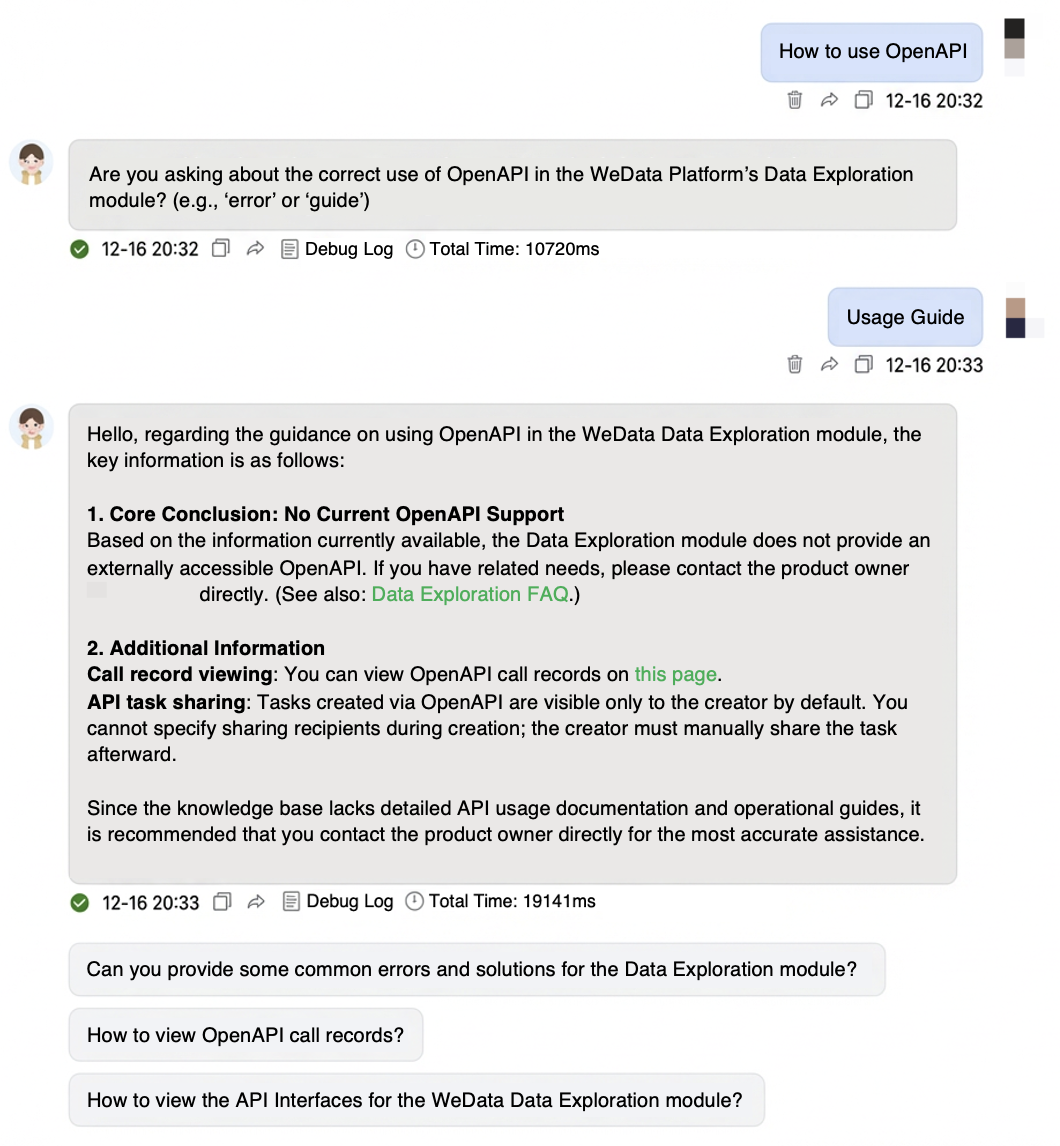}
    \caption{Product interface II: Chatbot.}
  \label{fig:chatbot_access}
\end{figure}

Our framework has been deployed in production on the company’s internal big-data platform. 
We provide two entry points (Figure~\ref{fig:in_console_diagnosis},~\ref{fig:chatbot_access}), targeting different usage scenarios. 
(1) In-console diagnosis: \sys is embedded into the big-data platform UI. When a job fails or behaves abnormally, users can trigger a one-click diagnosis directly from the execution page. 
(2) Chatbot access: \sys is integrated into the internal messaging tool as a chatbot, supporting not only error diagnosis but also consultation-style questions such as component syntax and usage. 

These two interfaces exhibit different strengths. 
The in-console diagnosis interface is tightly coupled with the big-data platform and can automatically populate the required inputs (e.g., failed query, runtime context, and error logs), enabling fast, first-line troubleshooting with minimal user effort.
However, it is mainly optimized for failure cases and is less flexible for open-ended inquiries. 
The chatbot interface supports flexible, multi-turn interaction and primarily serves ticket submitters and on-call engineers for both diagnosis and knowledge queries. 
Since conversational inputs are often incomplete, the service leverages the intent clarification module to ask follow-up questions and recover missing constraints. 

\subsection{Design Details}
In this part, we present the design of auxiliary components that complement the core ideas introduced in Section~\ref{sec:system_design}. 
These components mainly include intent clarification, retrieval, and summarization.
 
\textbf{Intent clarification and routing}. 
Since user requests in industrial settings are often underspecified, \sys performs intent completion through disambiguation and multi-turn follow-up questions. 
Given the request type, this component checks whether a minimal set of required fields is present (e.g., task ID, error logs). 
If critical inputs are missing (e.g., diagnosing a failure without any logs), \sys asks targeted questions and guides users to paste the most relevant log snippets for diagnosis. 
For usability, it also normalizes noisy inputs (e.g., trimming duplicated stack traces) before passing them downstream.

After the intent is completed, \sys extracts routing keywords from the clarified request and matches them against a predefined keyword set using cosine similarity. 
Based on the similarity scores, the request is dispatched to the most appropriate downstream agent or falls back to the general pipeline when no confident match is found.

\textbf{Retrieval}. For internal SOPs and domain knowledge, \sys adopts a hybrid retrieval strategy that combines lexical matching (token-based search) with embedding-based semantic similarity, which improves robustness to both exact error codes and paraphrased descriptions. 
Retrieved contents are indexed at the passage level and reranked jointly, after which the system selects the top candidates for the subsequent filtering and reasoning steps. 
For external knowledge, \sys uses the Google Search API to obtain relevant pages and the Jina API to fetch and convert web content into clean text that is suitable for downstream processing.
The results are further quality-filtered to reduce unreliable sources. 

\textbf{Summarization}. 
The summarization module generates the final response by synthesizing the clarified intent and the verified evidence accumulated during DeepSearch. 
It produces a structured output tailored to the request type: for diagnosis tasks, it reports the most likely root cause, supporting evidence (with citations to retrieved passages), and actionable investigation and resolution steps; for consultation queries, it provides a concise explanation and recommended usage patterns. 
To reduce hallucinations, the module is constrained to ground key claims in retrieved evidence, explicitly separates confirmed findings from hypotheses, and includes ``missing information'' tags when the evidence is still insufficient for a definitive conclusion.

Beyond the basic implementation, we incorporate several practical optimizations to improve reliability. 
First, \sys maintains a lightweight session memory so that follow-up questions can reuse previously provided context and avoid repeated queries.
Second, the system enforces safety and cost controls, including a search budget, and safe-response templates for high-risk actions. 
Finally, we log intermediate states (e.g., clarified intent, selected evidence, and final outputs) to enable post-hoc debugging. 
This is particularly useful when a request is escalated to a ticket, as the logs help locate failures in routing, retrieval, or tool execution.

\subsection{Implementation Details}
In this section, we describe the implementation details of \sys, including the prompt design of two core modules and the deployment settings.

The prompt used for the DeepSearch planner is as follows.
\begin{lstlisting}[] 
# Task Description
You are an expert in big data platforms. Your task 
is to perform root cause analysis and provide 
solutions for users' problems.

# Tools
<platform_tools_description>

# Retrievers
You can only use retrievers available in the
**current Level**. Start in Level 1.
## Level 1
<retrievers_description>
## Level 2
<retrievers_description>
...

# Workflow
If information is sufficient, return the final answer; 
Otherwise, analyze the issue and then choose one of 
the following actions:
(a) Call a platform tool;
(b) Retrieve knowledge: 
    - decide to stay at the current Level, or move to
      the next Level;
    - generate a query to retrieve.

# System Background Knowledge
<system_overview_description>
<system_component_knowledge>

# Input
query: <clarified_input>

# Final Answer Format
<output_format>

\end{lstlisting}











Due to privacy concerns around production logs in big data systems, \sys is implemented without closed-source third-party APIs. 
To support iterative reasoning with low latency, we deploy different models for different stages: we use Qwen3-30B-A3B~\cite{yang2025qwen3} for task planning and reflection to reduce end-to-end latency during multi-iteration search, while intent clarification and final summarization are handled by DeepSeek V3.2~\cite{liu2025deepseek} to improve response quality. 
Both models are deployed on NVIDIA H20 GPUs. 

%% file: experiment.tex
\section{Experiments}
\label{sec:exp}
To demonstrate the effectiveness of \sys, we aim to answer the following questions:
1) \sys has a clear online impact after deployment, as reflected by a reduced ticket escalation rate and lower overall ticket volume;
2) For specialized troubleshooting scenarios, \sys achieves higher answer accuracy and usefulness than representative SOTA baselines;
3) The overall system design is reasonable, and each major component contributes measurable improvements to end-to-end performance.

\subsection{Experiment Setups}

\quad\textbf{Dataset}.
To evaluate \sys on specialized troubleshooting scenarios, we build a benchmark from production incidents collected on the Tencent Big Data platform. 
Specifically, on-call engineers filtered out the 100 challenging ticket cases collected over five months.
Each case consists of the original user request, which may include user query, SQL statements, and log snippets. 
To obtain reliable ground-truth labels, we invited on-call engineers to annotate each case with a canonical root cause, which is used as the ground truth answer in our evaluation.
The dataset covers a range of frequent issues, including dependency incompatibility, OOM failures, SQL syntax errors, etc.
Moreover, 55\% of cases lack error logs or SQL statements to reflect the common lack of complete information in real-world scenarios.

\textbf{Baselines}.
For end-to-end comparison to validate the effectiveness of the specialized agents, we compare \sys with the following core techniques commonly used in representative SOTA LLM-based frameworks:
(1) CoT~\cite{wei2022chain}: A prompting-only baseline. The LLM is provided only with basic system background information without any external knowledge.
(2) RAG~\cite{singh2024panda, zhang2024automated, cuiaetherlog}: A single-turn retrieval-augmented baseline. The agent retrieves once from a flat knowledge base that contains all knowledge without reflection on retrieval results, and then directly generates the answer.
(3) Vanilla DeepSearch~\cite{zhang2024mabc, ren2025multi, pei2025flow}: A multi-turn retrieval-augmented baseline. The agent performs iterative retrieval across flat multiple specialized knowledge bases without reflection on retrieval results, and autonomously decides whether the accumulated evidence is sufficient to output the final answer.

\textbf{Evaluation Metric}.
To characterize the production impact after deployment, we report two online metrics: ticket escalation rate (i.e., the fraction of assistant interactions that are still escalated to an online ticket) and the change in overall ticket volume.
On the specialized troubleshooting benchmark, we evaluate answer quality using accuracy and usefulness, and report efficiency metrics including latency and the average number of retrieval iterations.
Usefulness refers to the proportion of cases where human experts judge the output to be actionable and reference-worthy for the analysis process, even if not fully correct.

\textbf{Knowledge Base}.
We maintain an SOP knowledge base with 679 entries to accommodate different user inputs, indexing both error-code signatures and error descriptions (for log-absent consultations where users only provide symptom narratives).
The domain-specific knowledge is organized into five knowledge bases, covering various parts of systems such as the SQL distribution platform, computing engine, and data storage, with a total of 538 entries.
The Qwen3-Embedding-8B model is used to embed the keys of these knowledge base entries.


\textbf{Experiment Settings}.
All LLMs are configured with a temperature of 0.1 and a top\_p of 0.95.
For RAG, we retrieve the top-10 entries. 
For other retrieval-based methods, we retrieve the top-5 entries per iteration.
Each sample is inferred only once (i.e., we do not perform multiple runs or majority voting).

\subsection{In-production Performance}
We have deployed \sys on the Tencent Big Data platform in production. 
We evaluate its online impact via the ticket escalation rate and the change in overall ticket volume.

\begin{figure}
    \centering
    \includegraphics[width=0.95\linewidth]{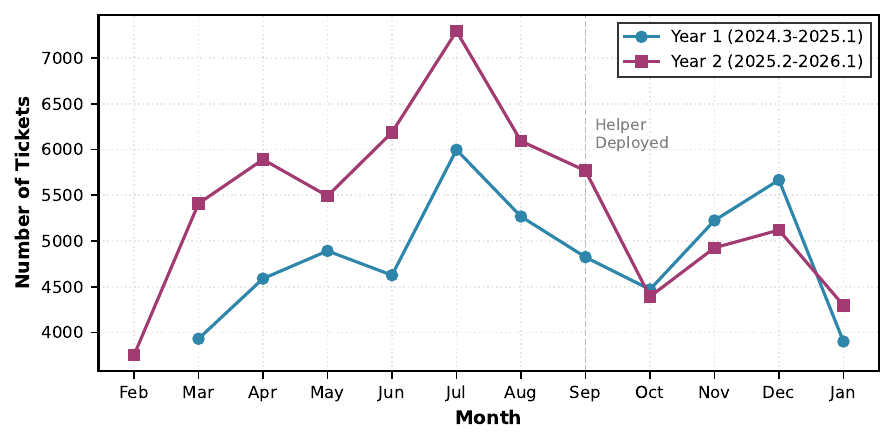}
    \caption{Ticket volume from Mar 2024 to Jan 2026.}
    \label{fig:ticket_volume}
\end{figure}

We collected raw user queries handled by \sys over 7 days.
Among these queries, 33.5\% were directly resolved without being escalated to tickets. 
The average end-to-end response latency is 15.05 seconds, which meets enterprise users' expectations for an interactive assistant.
Figure~\ref{fig:ticket_volume} shows the ticket volume trend from March 2024 (when the data pipeline became available) to January 2026. 
As the user base of the big data platform grew, we observe that the total ticket volume in 2025 increased by about 24.1\% year-over-year compared with 2024.
Since \sys was deployed in September 2025, ticket volume has shown a clear downward trend.
After accounting for the overall ticket growth, \sys reduces the total ticket volume by around 20.8\%.

\begin{table}[t]
    \centering
    \caption{Comparison of key techniques applied in SOTA analysis frameworks.}
    \resizebox{0.38\pdfpagewidth}{!}{
    \begin{tabular}{ccccc}
    \toprule
    & \multicolumn{2}{c}{Effectiveness(\%)} & \multicolumn{2}{c}{Latency(s)}\\
    & accuracy & usefulness & Avg & P90\\
    \midrule
    CoT & 54 & 62 & 25.91 & 42 \\
    RAG & 57 & 68 & 19.96 & 30 \\
    Vanilla DeepSearch & 62 & 71 & 19.93 & 31 \\
    \sys & 73 & 81 & 24.16 & 42 \\
    \bottomrule
    \end{tabular}
    }
    \label{tab:baseline}
\end{table}

\subsection{Comparison of Baselines}
For specialized troubleshooting scenarios, we compare \sys with three baselines: CoT, RAG, and vanilla DeepSearch, which represent the core techniques adopted by current SOTA analysis frameworks as introduced in experimental setups.
The results are shown in Table~\ref{tab:baseline}.

\textbf{Effectiveness}. 
\sys achieves the best overall diagnostic quality, with accuracy and usefulness reaching 73\% and 81\%. 
Vanilla DeepSearch ranks second (62\%/71\%), which supports the benefit of an iterative retrieval process. 
Concretely, iterative retrieval allows the system to refine search queries based on intermediate hypotheses and recover from an initially suboptimal retrieval result. 
This is particularly important for specialized tasks, where the key evidence is often scattered across different logs/docs and may not be surfaced by a single retrieval. 
In contrast, one-shot RAG is more sensitive to the first retrieved set and thus more likely to miss critical clues, resulting in lower accuracy/usefulness (57\%/68\%). 
CoT, without grounding evidence, further suffers from unsupported reasoning and produces the lowest effectiveness (54\%/62\%).

\textbf{Latency}. 
Compared with \sys, RAG is faster as it relies on single-round retrieval, with an average latency of 19.96 seconds.
Although vanilla DeepSearch may require multiple iterations, its average latency is comparable to RAG because it retrieves only half as many entries per iteration.
Due to the reflection step in our DeepSearch workflow, the average latency increases by about 4 seconds, reaching 24.16 seconds.
Interestingly, CoT shows relatively high latency (25.91 seconds), suggesting that without reliable external evidence the model tends to spend more time on free-form reasoning, whereas trusted knowledge sources can shorten the overall reasoning process.
Overall, \sys provides the best quality–latency trade-off among all compared methods.

\subsection{Ablation on Knowledge Base Design}
\label{sec:knowledge_analysis}
To understand how the knowledge base affects the overall performance of \sys, we conduct the following two experiments.

\begin{table}[t]
    \centering
    \caption{Comparison of different knowledge base structures. H and F denote hierarchical and flat structures, respectively.}
    \resizebox{0.38\pdfpagewidth}{!}{
    \begin{tabular}{cccccc}
    \toprule
    & \multicolumn{2}{c}{Effectiveness(\%)} & \multicolumn{2}{c}{Latency(s)} & Retrieval\\
    & accuracy & usefulness & Avg & P90 & iterations\\
    \midrule
    Vanilla DeepSearch (F) & 62 & 71 & 19.93 & 31 & 1.16\\
    Vanilla DeepSearch (H) & 65 & 73 & 19.18 & 34 & 1.26 \\
    \sys (F) & 67 & 76 & 27.78 & 47 & 1.63 \\
    \sys (H) & 73 & 81 & 24.16 & 42 & 1.30 \\
    \bottomrule
    \end{tabular}
    }
    \vspace{2mm}
    \label{tab:knowledge_base_structure}
\end{table}

\textbf{Knowledge Base Structure}.
We first evaluate the impact of knowledge base structure by comparing a flat knowledge base (Flat) against our proposed hierarchical knowledge base (Hierarchical). 
For each construction strategy, we compare \sys with a naive DeepSearch baseline that performs iterative retrieval without result reflection.

Our results show that both methods perform worse under the Flat setting. 
Specifically, compared with the Hierarchical setting, the answer accuracy and usefulness of \sys drop by 6\% and 5\% under the Flat setting, respectively. 
Meanwhile, Flat also leads to more retrieval iterations, increasing the number of retrieval rounds by 0.33. 
As a result, the average latency increases by 3.62 seconds. 
These results indicate that hierarchical structure not only improves answer quality but also reduces the retrieval cost by guiding the search toward higher-priority and more actionable knowledge.

\begin{table}[t]
    \centering
    \caption{Comparison of different knowledge base layers.}
    \resizebox{0.38\pdfpagewidth}{!}{
    \begin{tabular}{cccccc}
    \toprule
    & \multicolumn{2}{c}{Effectiveness(\%)} & \multicolumn{2}{c}{Latency(s)} & Retrieval\\
    & accuracy & usefulness & Avg & P90 & iterations\\
    \midrule
    \sys w/o SOP & 63 & 69 & 29.66 & 49 & 2.10  \\
    \sys w/o domain & 66 & 75 & 22.89 & 49 & 1.17 \\
    \sys w/o external & 71 & 77 & 24.11 & 40 & 1.34 \\
    \sys & 73 & 81 & 24.16 & 42 & 1.30 \\
    \bottomrule
    \end{tabular}
    }
    \vspace{2mm}
    \label{tab:knowledge_base_layer}
\end{table}

\textbf{Knowledge Base Layer}.
We further conduct a study to quantify the contribution of each layer in the hierarchical knowledge base by removing one layer at a time. 

Overall, removing any layer degrades the system, suggesting that each layer makes complementary contributions to the entire pipeline. 
The external knowledge layer has the smallest impact on both accuracy and usefulness (-2\%/-4\%), because most real-world tasks can already be handled effectively using SOPs and domain-specific knowledge. 
In contrast, removing the \textbf{SOP layer} leads to the most evident quality drop and latency regression: the accuracy reduces by 10\% while the average retrieval iterations increase by 0.80. 
This is because SOPs typically provide high-precision, workflow-oriented guidance that quickly narrows down the search space and reduces unnecessary retrieval iterations.

Interestingly, removing the domain-specific knowledge layer decreases average latency by 0.73 seconds.
We attribute this behavior to the fact that internal documents are often broad and loosely structured, which means a thorough retrieval process may trigger multiple searches over this layer before converging. 
Removing this layer reduces the retrieval cost and thus shortens latency, but it leads to a large drop in answer quality (-7\%/-6\%).

\subsection{Ablation on System Details}
We further investigate how specific design choices affect \sys by conducting ablation studies on (i) the DeepSearch workflow and (ii) the backbone model. 
These experiments aim to characterize the quality--latency trade-off in practical deployment.

\textbf{DeepSearch Process}.
We further evaluate the workflow design in the DeepSearch engine, focusing on result reflection introduced in \sys. 
Compared with a naive DeepSearch pipeline, our DeepSearch adds a result reflection stage that filters out irrelevant evidence before summarization, aiming to reduce hallucinations and wrong attributions.

Table~\ref{tab:knowledge_base_structure} shows the results of \sys and the naive DeepSearch baseline under the same knowledge base structure. 
We observe that our method, which includes the reflection-based filtering, has more retrieval iterations than the naive DeepSearch baseline, while our method increases accuracy and usefulness both by +8\% but incurs higher latency (24.16 vs. 19.18 seconds). 
We attribute the difference to the reflection-based filtering. 
Concretely, the naive DeepSearch baseline is slightly faster because it skips the reflection step, but this speedup comes at the cost of more hallucination and lower answer quality. 
As online users typically expect responses within 30 seconds and \sys meets this latency requirement, we adopt the more accurate workflow in practice.

\textbf{Backbone Model}.
We also evaluate \sys under different backbone models. 
In addition to the default setting, we replace the Planner and the Filter with a larger model, Qwen-235B-A22B, while keeping the rest unchanged. 
This comparison helps quantify how model capacity affects the overall quality–latency trade-off.

Overall, using a larger model brings slight gains in diagnosis quality. 
We observe improvements in both accuracy and usefulness by 1\%. 
The larger model performs well in generating more complete investigation plans, and performing stricter evidence filtering, which reduces incorrect attributions in ambiguous cases.
However, these gains come with an increase in end-to-end latency. The average latency increases by 41.42 seconds (and P90 latency by 60 seconds), mainly due to higher per-call inference cost and longer intermediate reasoning traces produced by the larger model.
Considering user requirements in production, we use the smaller model as the backbone model, and reserve the larger-model configuration for cases where higher accuracy is preferred over latency.

\begin{table}[t]
    \centering
    \caption{Comparison of different backbone models.}
    \resizebox{0.38\pdfpagewidth}{!}{
    \begin{tabular}{ccccccc}
    \toprule
    & \multicolumn{2}{c}{Effectiveness(\%)} & \multicolumn{2}{c}{Latency(s)} & Retrieval\\
    & accuracy & usefulness & Avg & P90 & iterations\\
    \midrule
    \sys (30B) & 73 & 81 & 24.16 & 42 & 1.30 \\
    \sys (235B) & 74 & 82 & 65.58 & 102 & 3.00 \\
    \bottomrule
    \end{tabular}
    }
    \vspace{2mm}
    \label{tab:backbone_model}
\end{table}

\subsection{SOP Case Study}
\label{sec:sop_case_study}

We present a case study to illustrate how to extract SOP knowledge from historical tickets, following the pipeline in Section~\ref{sec:knowledge_self_enrichment}. 
The original ticket contains multi-turn conversation between users and on-call engineers as follows:

\begin{lstlisting}[]
# Ticket Conversation
OCE1: Hello, please describe your issue.
user1:  
  java.lang.NumberFormatException: For input string: 
  "yyyy-mm-dd hh:mm:ss"
  ...
user1: I don't quite understand the cause of this 
  error. The fields I queried don't have any date 
  fields in this format.
OCE1: You can check which field this data comes 
  from-maybe its type doesn't match the target field 
  type and it can't be written.
user1: There's the SQL.
...
OCE2: last_modified_time  bigint.
user1: @user2 Here.
...
user2: I see the editor only allows renaming. Can I 
  rename the old field to something else, and then 
  add a new last_modified_time field?
OCE1: Yes.
OCE1: @user1 Do you have any other issues that need 
  further assistance?
user1: Thanks everyone, it's working normally now.
OCE1: Okay.
\end{lstlisting}

Directly storing raw ticket dialogues is suboptimal for future reuse because actionable knowledge is intertwined with conversational noise, making it difficult to retrieve, compare, and reuse.
Therefore, we first use the SOP Author and Editor agent to convert the ticket into a structured SOP, including a normalized problem description, the core candidate root cause, and the corresponding investigation and resolution steps.

\begin{lstlisting}[]
# Structured SOP
{
  "problem_desc": "java.lang.NumberFormatException: 
  For input string: 'xxx'",
  "content": [
    {
      "root_cause": "Column type in metadata does not 
      match the actual stored data type.",
      "investigation_steps": [
        {
          "step": "1",
          "target": "Column schema",
          "action": "Compare the declared column type 
          with the real stored values",
          "observations": "..."
        }
      ],
      "resolution_steps": [
        {
          "step": "1",
          "action": "..."
        },
        ...
      ]
    }
  ]
}
\end{lstlisting}

Furthermore, to mitigate hallucination risk during SOP generation, we repeat the SOP generation process three times, and then use the SOP Reviewer agent to evaluate consistency across versions.
We define a stability score that equals the number of runs whose root-cause semantics agree.
The following example demonstrates the SOP Reviewer’s behavior: multiple candidate SOPs have semantically identical root causes, yielding a high stability score.

\begin{lstlisting}[]
# SOP Review Process
(1) Candidate SOPs Root Cause Analysis:
- Root cause in SOP 1: Column type in metadata does 
not match the actual stored data type.
- Root cause in SOP 2: Schema mismatch ...
- Root cause in SOP 3: ...

Above Root Causes are semantically identical.
Stability score is: 3.

(2) Final SOP:
{
  "problem_desc": "java.lang.NumberFormatException: 
  For input string: 'xxx'",
  "content": [
    ...
  ]
}
\end{lstlisting}

The example below illustrates a case where retrieval finds an older SOP with a similar problem description.
Rather than creating a new entry or discarding the old one, the SOP Curator updates the existing SOP entry by merging the new root cause as an additional branch when it is not covered by the old SOP.
As a result, the merged SOP supports multiple root cause branches for the same symptom, helping future operators quickly tell the difference between (i) a schema mismatch and (ii) unexpected characters in the data, both of which can lead to a NumberFormatException error.

\begin{lstlisting}[]
# Retrieved Similar SOP
{
  "problem_desc": "Caused by: java.lang.NumberFormat-
  Exception: For input string: \"xxx\"\\n ...",
  "content": [
    {
      "root_cause": "The data contains unparsable 
      characters, such as line breaks (\\n).",
      ...
    }
  ]
}

# Merged SOP
{
  "problem_desc": "java.lang.NumberFormatException: 
  For input string: 'xxx'",
  "content": [
    {
      "root_cause": "Column type in metadata does not 
      match the actual stored data type.",
      ...
    },
    {
      "root_cause": "The data contains unparsable 
      characters, such as line breaks (\\n).",
      ...
    }
  ]
  
}

\end{lstlisting}

%% file: relatedwork.tex
\section{Related Work}
\label{sec:related}
Modern big data platforms consist of diverse components (e.g., compute engines, storage, and SQL gateways) and evolve rapidly. 
As a result, users may submit diverse requests ranging from failure reports and error diagnosis to operational consultations, and handling these requests is costly for both users and on-call engineers.
This has motivated the development of AIOps systems that aim to automate root-cause analysis (RCA) and user-facing diagnosis.
Existing AIOps solutions for big data platforms can be grouped into three categories:

(1) Rule-based: 
Traditional AIOps solutions are primarily designed for system-centric monitoring rather than open-ended user support.  
They typically detect anomalies from statistical metrics and perform RCA by applying predefined rules to metric signals to identify anomalies most likely responsible for the observed performance degradation, as exemplified by BigRoots~\cite{zhou2018bigroots}, and AutoDiagn~\cite{demirbaga2021autodiagn}.
Such rules are largely handcrafted based on domain expertise, which limits scalability and leaves many real-world failure modes uncovered.

(2) Machine learning-based: 
Similar to rule-based approaches, machine learning-based methods are typically developed for specific scenarios and address failures through anomaly detection followed by RCA.
Representative systems train machine learning models (e.g., classifiers or regressors) on historical data to capture recurring anomaly patterns, such as LADRA~\cite{lu2019ladra}, Rootpath~\cite{demirbaga2023rootpath} and AUTOMATE~\cite{liu2024automate}. 
Despite improved adaptability, these methods typically depend on carefully engineered features and labeled training data, making them costly to maintain under rapid platform evolution.
In addition, their outputs are often restricted to a predefined set of coarse-grained root-cause categories, which limits their ability to generate fine-grained, actionable guidance for users.

(3) LLM-based: 
Recent works have explored the use of LLMs to build AIOps systems. 
Compared with rule-based or traditional machine-learning approaches, LLMs offer stronger flexibility in interpreting diverse textual inputs.
However, deploying LLMs in production AIOps remains challenging: effective systems typically require domain knowledge and carefully designed workflows to improve reliability and mitigate hallucinations.
Existing LLM-based AIOps systems span a spectrum of architectures, including single-agent designs~\cite{singh2024panda, cuiaetherlog} and 
multi-agent frameworks~\cite{zhou2024llm, zhou2025gaussmaster, zhou2023d}.
Among them, AetherLog~\cite{cuiaetherlog} and D-Bot~\cite{zhou2023d} are mainly log-driven, taking failure logs or structured anomaly descriptions as input, rather than supporting open-ended, user-initiated queries. 
In contrast, LLMDB~\cite{zhou2024llm}, Panda~\cite{singh2024panda} and GaussMaster~\cite{zhou2025gaussmaster} build a conversational assistant with intent classification and clarification.
Moreover, despite leveraging various knowledge sources, most systems organize them as a flat corpus without differentiating their relative importance or reliability.
In addition, knowledge evolution is underexplored.
Most prior work relies on offline knowledge construction~\cite{zhou2024llm, zhou2025gaussmaster, zhou2023d} or simple knowledge maintenance, lacking robust quality control~\cite{singh2024panda, cuiaetherlog}.


Beyond studies in big data systems, LLM-based AIOps has also been studied extensively in cloud and microservice environments.
Prior work includes systems built on simple RAG-based pipelines~\cite{zhang2024automated} as well as a larger body of approaches developing interactive ReAct paradigms~\cite{roy2024exploring, wang2024rcagent, zhang2024mabc, ren2025multi, pei2025flow}.
However, many of these systems are primarily incident-driven: they are triggered by detected alerts or incidents with fixed input schemas, which limits their ability to handle open-ended, user-initiated questions. 
In addition, knowledge self-enrichment is not a central focus in most existing designs, and several works simply integrate successful historical diagnosis traces into their knowledge bases~\cite{ren2025multi, pei2025flow}, without systematic mechanisms for filtering, summarization, and governance of newly introduced knowledge.
Finally, only Flow-of-Action~\cite{pei2025flow} explicitly emphasizes prioritizing SOPs over historical incident records. 
More generally, the importance of differentiating knowledge types and organizing them in a hierarchical, priority-aware manner remains insufficiently explored.

%% file: reference.bib
@article{cuapusneanu2025reshaping,
  title={Reshaping the Digital Economy with Big Data: A Meta-Analysis of Trends and Technological Evolution},
  author={C{\u{a}}pușneanu, Sorinel and Barbu, Cristian-Marian and Solomon, Alina-Georgiana and Rakos, Ileana-Sorina},
  journal={Electronics},
  volume={14},
  number={13},
  pages={2709},
  year={2025},
  publisher={MDPI}
}

@article{shahnawaz2025comprehensive,
  title={A comprehensive survey on big data analytics: Characteristics, tools and techniques},
  author={Shahnawaz, Mohammad and Kumar, Manish},
  journal={ACM Computing Surveys},
  volume={57},
  number={8},
  pages={1--33},
  year={2025},
  publisher={ACM New York, NY}
}

@article{rahaman2024evaluating,
  title={Evaluating SQL understanding in large language models},
  author={Rahaman, Ananya and Zheng, Anny and Milani, Mostafa and Chiang, Fei and Pottinger, Rachel},
  journal={arXiv preprint arXiv:2410.10680},
  year={2024}
}

@article{crupi2025effectiveness,
  title={On the Effectiveness of LLM-as-a-judge for Code Generation and Summarization},
  author={Crupi, Giuseppe and Tufano, Rosalia and Velasco, Alejandro and Mastropaolo, Antonio and Poshyvanyk, Denys and Bavota, Gabriele},
  journal={IEEE Transactions on Software Engineering},
  year={2025},
  publisher={IEEE}
}

@article{huynh2025large,
  title={Large language models for code generation: A comprehensive survey of challenges, techniques, evaluation, and applications},
  author={Huynh, Nam and Lin, Beiyu},
  journal={arXiv preprint arXiv:2503.01245},
  year={2025}
}

@article{guo2025deepseek,
  title={DeepSeek-R1 incentivizes reasoning in LLMs through reinforcement learning},
  author={Guo, Daya and Yang, Dejian and Zhang, Haowei and Song, Junxiao and Wang, Peiyi and Zhu, Qihao and Xu, Runxin and Zhang, Ruoyu and Ma, Shirong and Bi, Xiao and others},
  journal={Nature},
  volume={645},
  number={8081},
  pages={633--638},
  year={2025},
  publisher={Nature Publishing Group UK London}
}

@article{cheng2025empowering,
  title={Empowering llms with logical reasoning: A comprehensive survey},
  author={Cheng, Fengxiang and Li, Haoxuan and Liu, Fenrong and Van Rooij, Robert and Zhang, Kun and Lin, Zhouchen},
  journal={arXiv preprint arXiv:2502.15652},
  year={2025}
}

@article{chen2025towards,
  title={Towards reasoning era: A survey of long chain-of-thought for reasoning large language models},
  author={Chen, Qiguang and Qin, Libo and Liu, Jinhao and Peng, Dengyun and Guan, Jiannan and Wang, Peng and Hu, Mengkang and Zhou, Yuhang and Gao, Te and Che, Wanxiang},
  journal={arXiv preprint arXiv:2503.09567},
  year={2025}
}

@article{lewis2020retrieval,
  title={Retrieval-augmented generation for knowledge-intensive nlp tasks},
  author={Lewis, Patrick and Perez, Ethan and Piktus, Aleksandra and Petroni, Fabio and Karpukhin, Vladimir and Goyal, Naman and K{\"u}ttler, Heinrich and Lewis, Mike and Yih, Wen-tau and Rockt{\"a}schel, Tim and others},
  journal={Advances in neural information processing systems},
  volume={33},
  pages={9459--9474},
  year={2020}
}

@inproceedings{shao2023enhancing,
  title={Enhancing retrieval-augmented large language models with iterative retrieval-generation synergy},
  author={Shao, Zhihong and Gong, Yeyun and Shen, Yelong and Huang, Minlie and Duan, Nan and Chen, Weizhu},
  booktitle={Findings of the Association for Computational Linguistics: EMNLP 2023},
  pages={9248--9274},
  year={2023}
}

@article{singh2025agentic,
  title={Agentic retrieval-augmented generation: A survey on agentic rag},
  author={Singh, Aditi and Ehtesham, Abul and Kumar, Saket and Khoei, Tala Talaei},
  journal={arXiv preprint arXiv:2501.09136},
  year={2025}
}

@inproceedings{fan2024survey,
  title={A survey on rag meeting llms: Towards retrieval-augmented large language models},
  author={Fan, Wenqi and Ding, Yujuan and Ning, Liangbo and Wang, Shijie and Li, Hengyun and Yin, Dawei and Chua, Tat-Seng and Li, Qing},
  booktitle={Proceedings of the 30th ACM SIGKDD conference on knowledge discovery and data mining},
  pages={6491--6501},
  year={2024}
}

@article{xi2025survey,
  title={A survey of llm-based deep search agents: Paradigm, optimization, evaluation, and challenges},
  author={Xi, Yunjia and Lin, Jianghao and Xiao, Yongzhao and Zhou, Zheli and Shan, Rong and Gao, Te and Zhu, Jiachen and Liu, Weiwen and Yu, Yong and Zhang, Weinan},
  journal={arXiv preprint arXiv:2508.05668},
  year={2025}
}

@article{wu2025webdancer,
  title={Webdancer: Towards autonomous information seeking agency},
  author={Wu, Jialong and Li, Baixuan and Fang, Runnan and Yin, Wenbiao and Zhang, Liwen and Tao, Zhengwei and Zhang, Dingchu and Xi, Zekun and Fu, Gang and Jiang, Yong and others},
  journal={arXiv preprint arXiv:2505.22648},
  year={2025}
}

@misc{openai2025gpt52,
  author = {{OpenAI}},
  title = {Introducing Deep Research},
  url = {https://openai.com/zh-Hans-CN/index/introducing-deep-research/},
  year = {2025},
}

@misc{google2025geminideepresearch,
  author = {{Google}},
  title = {Try Deep Research and our new experimental model in Gemini, your AI assistant},
  url = {https://blog.google/
products/gemini/google-gemini-deep-research/},
  year = {2024}
}

@article{zhou2018bigroots,
  title={Bigroots: An effective approach for root-cause analysis of stragglers in big data system},
  author={Zhou, Honggang and Li, Yunchun and Yang, Hailong and Jia, Jie and Li, Wei},
  journal={IEEE Access},
  volume={6},
  pages={41966--41977},
  year={2018},
  publisher={IEEE}
}

@article{demirbaga2021autodiagn,
  title={Autodiagn: An automated real-time diagnosis framework for big data systems},
  author={Demirbaga, Umit and Wen, Zhenyu and Noor, Ayman and Mitra, Karan and Alwasel, Khaled and Garg, Saurabh and Zomaya, Albert Y and Ranjan, Rajiv},
  journal={IEEE Transactions on Computers},
  volume={71},
  number={5},
  pages={1035--1048},
  year={2021},
  publisher={IEEE}
}

@inproceedings{liu2024automate,
  title={AUTOMATE: Automatic anomaly detection and root cause analysis framework for hadoop},
  author={Liu, Xinyuan and Jha, Devki Nandan and Li, Yinhao and Barika, Mutaz and Demirbaga, Umit and Ranjan, Rajiv},
  booktitle={2024 International Conference on Meta Computing (ICMC)},
  pages={213--222},
  year={2024},
  organization={IEEE}
}

@article{lu2019ladra,
  title={LADRA: Log-based abnormal task detection and root-cause analysis in big data processing with Spark},
  author={Lu, Siyang and Wei, Xiang and Rao, Bingbing and Tak, Byungchul and Wang, Long and Wang, Liqiang},
  journal={Future Generation Computer Systems},
  volume={95},
  pages={392--403},
  year={2019},
  publisher={Elsevier}
}

@article{demirbaga2023rootpath,
  title={Rootpath: Root cause and critical path analysis to ensure sustainable and resilient consumer-centric big data processing under fault scenarios},
  author={Demirbaga, Umit and Aujla, Gagangeet Singh},
  journal={IEEE Transactions on Consumer Electronics},
  volume={70},
  number={1},
  pages={1493--1500},
  year={2023},
  publisher={IEEE}
}

@article{singh2024panda,
  title={Panda: Performance debugging for databases using LLM agents.(2024)},
  author={Singh, Vikramank and Vaidya, Kapil Eknath and Kumar, Vinayshekhar Bannihatti and Khosla, Sopan and Narayanaswamy, Murali and Gangadharaiah, Rashmi and Kraska, Tim},
  year={2024}
}

@article{zhou2025gaussmaster,
  title={GaussMaster: An LLM-based Database Copilot System},
  author={Zhou, Wei and Sun, Ji and Zhou, Xuanhe and Li, Guoliang and Liu, Luyang and Wu, Hao and Wang, Tianyuan},
  journal={arXiv preprint arXiv:2506.23322},
  year={2025}
}

@article{zhou2024llm,
  title={Llm-enhanced data management},
  author={Zhou, Xuanhe and Zhao, Xinyang and Li, Guoliang},
  journal={arXiv preprint arXiv:2402.02643},
  year={2024}
}

@article{zhou2023d,
  title={D-bot: Database diagnosis system using large language models},
  author={Zhou, Xuanhe and Li, Guoliang and Sun, Zhaoyan and Liu, Zhiyuan and Chen, Weize and Wu, Jianming and Liu, Jiesi and Feng, Ruohang and Zeng, Guoyang},
  journal={arXiv preprint arXiv:2312.01454},
  year={2023}
}

@article{wei2022chain,
  title={Chain-of-thought prompting elicits reasoning in large language models},
  author={Wei, Jason and Wang, Xuezhi and Schuurmans, Dale and Bosma, Maarten and Xia, Fei and Chi, Ed and Le, Quoc V and Zhou, Denny and others},
  journal={Advances in neural information processing systems},
  volume={35},
  pages={24824--24837},
  year={2022}
}

@inproceedings{zhang2024automated,
  title={Automated root causing of cloud incidents using in-context learning with GPT-4},
  author={Zhang, Xuchao and Ghosh, Supriyo and Bansal, Chetan and Wang, Rujia and Ma, Minghua and Kang, Yu and Rajmohan, Saravan},
  booktitle={Companion Proceedings of the 32nd ACM International Conference on the Foundations of Software Engineering},
  pages={266--277},
  year={2024}
}

@inproceedings{roy2024exploring,
  title={Exploring llm-based agents for root cause analysis},
  author={Roy, Devjeet and Zhang, Xuchao and Bhave, Rashi and Bansal, Chetan and Las-Casas, Pedro and Fonseca, Rodrigo and Rajmohan, Saravan},
  booktitle={Companion proceedings of the 32nd ACM international conference on the foundations of software engineering},
  pages={208--219},
  year={2024}
}

@inproceedings{pei2025flow,
  title={Flow-of-Action: SOP Enhanced LLM-Based Multi-Agent System for Root Cause Analysis},
  author={Pei, Changhua and Wang, Zexin and Liu, Fengrui and Li, Zeyan and Liu, Yang and He, Xiao and Kang, Rong and Zhang, Tieying and Chen, Jianjun and Li, Jianhui and others},
  booktitle={Companion Proceedings of the ACM on Web Conference 2025},
  pages={422--431},
  year={2025}
}

@inproceedings{wang2024rcagent,
  title={Rcagent: Cloud root cause analysis by autonomous agents with tool-augmented large language models},
  author={Wang, Zefan and Liu, Zichuan and Zhang, Yingying and Zhong, Aoxiao and Wang, Jihong and Yin, Fengbin and Fan, Lunting and Wu, Lingfei and Wen, Qingsong},
  booktitle={Proceedings of the 33rd ACM International Conference on Information and Knowledge Management},
  pages={4966--4974},
  year={2024}
}

@article{ren2025multi,
  title={The Multi-Agent Fault Localization System Based on Monte Carlo Tree Search Approach},
  author={Ren, Rui},
  journal={arXiv preprint arXiv:2507.22800},
  year={2025}
}

@inproceedings{zhang2024mabc,
  title={mABC: multi-Agent Blockchain-Inspired Collaboration for root cause analysis in micro-services architecture},
  author={Zhang, Wei and Guo, Hongcheng and Yang, Jian and Tian, Zhoujin and Zhang, Yi and Chaoran, Yan and Li, Zhoujun and Li, Tongliang and Shi, Xu and Zheng, Liangfan and others},
  booktitle={Findings of the Association for Computational Linguistics: EMNLP 2024},
  pages={4017--4033},
  year={2024}
}

@INPROCEEDINGS{cuiaetherlog,
  author={Cui, Tianyu and Fu, Ruowei and Liu, Changchang and Ji, Yuhe and Gu, Wenwei and Zhang, Shenglin and Sun, Yongqian and Pei, Dan},
  booktitle={2025 IEEE 36th International Symposium on Software Reliability Engineering (ISSRE)}, 
  title={AetherLog: Log-based Root Cause Analysis by Integrating Large Language Models with Knowledge Graphs}, 
  year={2025},
  volume={},
  number={},
  pages={49-60},
  doi={10.1109/ISSRE66568.2025.00019}}

@article{yang2025qwen3,
  title={Qwen3 technical report},
  author={Yang, An and Li, Anfeng and Yang, Baosong and Zhang, Beichen and Hui, Binyuan and Zheng, Bo and Yu, Bowen and Gao, Chang and Huang, Chengen and Lv, Chenxu and others},
  journal={arXiv preprint arXiv:2505.09388},
  year={2025}
}

@article{liu2025deepseek,
  title={Deepseek-v3. 2: Pushing the frontier of open large language models},
  author={Liu, Aixin and Mei, Aoxue and Lin, Bangcai and Xue, Bing and Wang, Bingxuan and Xu, Bingzheng and Wu, Bochao and Zhang, Bowei and Lin, Chaofan and Dong, Chen and others},
  journal={arXiv preprint arXiv:2512.02556},
  year={2025}
}
